\documentclass[preprint2]{aastex}

\usepackage{graphicx}
\usepackage{epstopdf}
\usepackage{epsfig}
\shorttitle{Improved Photometry for DASCH} \shortauthors{Tang et al.}

\begin{document}


\title{Improved Photometry for the DASCH Pipeline} 


\author{Sumin Tang\altaffilmark{1, 2, 3}, Jonathan Grindlay\altaffilmark{1},
Edward Los\altaffilmark{1}, Mathieu Servillat\altaffilmark{1,4}}
\altaffiltext{1}{Harvard-Smithsonian Center for Astrophysics, 60
Garden St, Cambridge, MA 02138}
\altaffiltext{2}{Kavli Institute for Theoretical Physics, University of California, Santa Barbara, CA 93106}
\altaffiltext{3}{Division of Physics, Mathematics, \& Astronomy, California Institute of Technology, Pasadena, CA 91125, USA}
\altaffiltext{4}{Laboratoire AIM (CEA/DSM/IRFU/SAp, CNRS, Universite Paris Diderot),
 CEA Saclay, Bat. 709, 91191 Gif-sur-Yvette, France}


\begin{abstract}
The Digital Access to a Sky Century@Harvard (DASCH) project is digitizing the $\sim$500,000
glass plate images obtained (full sky) by the Harvard College Observatory from $1885-1992$.
Astrometry and photometry for each resolved object are derived with photometric rms values of $\sim$0.15mag
for the initial photometry analysis pipeline.
Here we describe new developments for DASCH photometry, applied to the Kepler field,
that has yielded further improvements, including better identification of image blends
and plate defects by measuring image profiles and astrometric deviations.
A local calibration procedure using nearby stars in a similar magnitude range
as the program star (similar to what has been done for visual photometry from the plates)
yields additional improvement for a net photometric rms $\sim$0.1mag.
We also describe statistical measures of light curves that are now used
in the DASCH pipeline processing to autonomously identify new variables.
The DASCH photometry methods described here are used in the pipeline processing
for the Data Releases of DASCH data (Grindlay et al. 2012; \textit{http://dasch.rc.fas.harvard.edu}),
as well as for the long-term variables discovered by DASCH in the Kepler field (Tang et al, in preparation).
\end {abstract}


\section{Introduction}
The Digital Access to a Sky Century@Harvard (DASCH) is an ongoing project to digitize and analyze
the scientific data contained in half a million astrophotographic plates that cover both the
northern and southern skies from the 1880s to the 1980s (Grindlay et al. 2009, 2012).
During the development phase for the high speed DASCH scanner
(Simcoe et al. 2006) and analysis software, we have scanned
over $20,000$ plates in five selected fields covering M44, 3C 273, Baade's Window,
the Kepler field, and the Large Magellanic Cloud (LMC).
The initial photometry and astrometry developments are described in Laycock et al. (2010; hereafter L10).
The pipeline and database is described in Los et al. (2011).
Further astrometry development is described in Servillat et al. (2011).
As a pilot scientific study,
we have discovered interesting long-term variables with $\sim1$ mag variations over years to decades,
which provide important information about dust processes, accretion physics
and possible nuclear shell burning on the surface of white dwarf (Tang et al. 2010, 2011, 2012).

Observational astronomy has now entered an unprecedented era of time domain astrophysics (TDA).
The Palomar Transient Factory (PTF; Law et al. 2009), the Panoramic Survey Telescope and Rapid Response System
(Pan-STARRS; Kaiser et al. 2002), and the Catalina Real-Time Transient Survey (CRTS;
Djorgovski et al. 2011), are all steps toward the future Large Synoptic Survey Telescope (LSST; Walker et al. 2003).
DASCH, with its unique 100 year time coverage and sampling,
opens a new window for studying variables, especially those with long timescale variations.
This requires the correct removal of plate defects (dust, scratches, etc. that are more likely on old plate images)
as well as good astrometry and photometry on each resolved object.

Here we describe further photometry developments beyond L10,
including identifying blended images,
filtering out dubious detections using image profiles and large astrometric errors,
and improved local calibration using neighbors with similar magnitudes.
These developments are all essential to finding real variables on DASCH plates.
We also describe statistical measurements of light curves, which will be used for selecting variables.
A subsequent paper in preparation will report on
the details of the initial DASCH variable search for the Kepler field (Tang et al. in prep.).

\section{The Initial Pipeline and Plates in the Kepler field}

Here we briefly summarize the initial photometric pipeline.
More details can be found at L10.
The DASCH scanner outputs 12 bit per pixel as a direct measure of the plate transmitted intensity, which is then inverted to give a positive image.
Note that the images are in photographic densities, which are non-linear, and are different from linear flux.
The bias level is about 50 ADU, and the maximum is 4095 ADU.
We divided each plates into 9 annular bins,
and fit the calibration curve in each bin separately,
to correct vignetting.
After extensive experiment, isophotal magnitude from SExtractor is adopted as the instrumental magnitude,
which has better performance than area or aperture magnitudes.
A bright star can cover $>1000$ pixels, providing a large range of isophotal magnitudes,
and therefore saturation is not a severe problem in most plates (see e.g. Fig. 2 in L10).
Most Harvard plates are blue sensitive without filters,
with a color response close to Johnson B.
However, a small fraction of the plates used red and yellow filters.
In order to generate consistent and precise magnitudes,
it is essential to understand the color response in each plate and correct it.
We derive the color term $C$, by minimizing rms in the calibration curve, as shown in Fig. 10 in L10.
The distributions of $C$ by plate series and year in the M44 field are shown in Fig. 11 in L10.

A particularly interesting field for long-term variability studies with DASCH is the
$\sim 10^{\circ} \times 10^{\circ}$ field observed by the Kepler mission (Borucki et al. 2010)
which provides very high sensitivity measures of variability on short timescales.
We have scanned $3735$ plates taken from 1885 to 1990, in or covering part of the Kepler field.
These plates are from 17 different series,
each typically represents a single telescope
(except the mb series which consists of data from multiple telescopes).
Table 1 lists information on number of plates in each series, telescope aperture,
field of view (FOV), median and RMS of limiting magnitude of the plates used in this work.
Note that each plate is divided into 9 annular bins for photometric calibration to
correct vignetting (as well as further corrections in ``local bins" for plate and sky variation corrections -- see L10),
and each annular bin has its own limiting magnitude value.
Most plates are blue-sensitive, close to Johnson B.
Since the only plates scanned for the ``Kepler field''
were chosen to contain part or all of the Kepler field,
the coverage for any given star decreases with increasing radius from the center of the Kepler field, as shown in Figure 1.
At any given point within the Kepler field of view (FOV),
there are at least 1500, 1000, 450, and 130 plates
down to $B=$12, 13, 14, and 15 magnitudes, respectively.
The Kepler field was, unfortunately, not observed as much by the more sensitive Harvard plates
(e.g. MC or A series) which cover many other fields, so these magnitude limits do not reflect the entire DASCH database.

For photometric calibration, we used the Kepler Input Catalog (KIC)
which contains Sloan-like griz photometry (Brown et al. 2011).
KIC covers $\sim177$ square degrees centered at RA 19:22:40 and Dec +44:30,
which is smaller than our scanned region.
In other fields, we used the GSC2.3 catalog (Lasker et al. 1990),
and the newly released AAVSO APASS catalog (Henden et al. 2012; \texttt{http://www.aavso.org/apass}) for photometric calibration
(the APASS catalog will be the default calibration for production DASCH data releases).
In this paper, we focus on the region covered by KIC.
DASCH light curves of Kepler planet host stars are presented in a companion paper, Tang et al. (2013).

\begin{figure}   [tb]
\epsfig{file=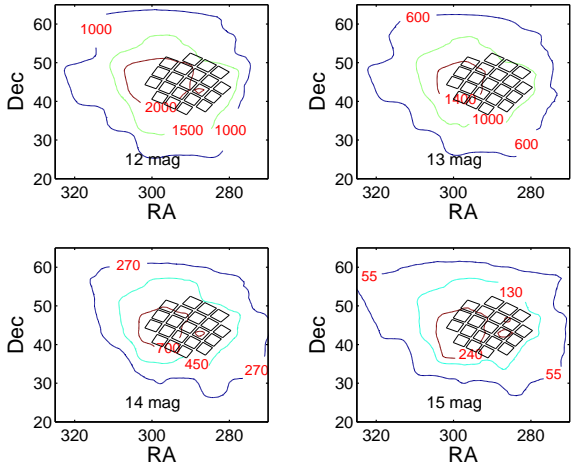, angle=0, width=\linewidth}
\caption{Coverage contour of number of digitized plates in or covering part of the Kepler FOV.
Four panels show number of plates reaching limiting magnitude of 12 mag (upper-left panel),
13 mag (upper-right panel), 14 mag (lower-left), and 15 mag (lower-right), respectively.
The Kepler FOV, as delineated by the CCD arrays, is overlaid (black squares) in each panel.}
\end{figure}

\section{Identify blended images}
With Galactic latitude of b $\sim13.5\pm8$ deg, the Kepler field is a moderately crowded field.
Moreover, due to overexposure and trailing on some plates, many bright objects may have
their photometry contaminated by nearby objects.
Plates are nonlinear detectors, and these blended images must be recognized
and removed (for now; image subtraction techniques will be explored for future applications) from the photometry analysis.
Otherwise, large errors/deviations from the true magnitude in the
blended stars will lead to fake variables, which will
pollute the pool of variables we are ultimately looking for.

SExtractor is used for object detection and photometry (Bertin \& Arnaults 1996).
It does some image deblending but requires a saddle point in the point
spread function (psf) that is usually lost for saturated stars.
Therefore, many blended images are not recognized as blended in SExtractor.
To overcome this problem, we use a second criterium to flag blended stars besides the SExtractor flag.
We adopt a critical separation radius, $\Delta r_{crit}$,
which is the sum of one half of the SExtractor full width half maximum (FWHM) radius
and the astrometric error.
For any SExtractor object, we search the input catalog (which is KIC in this paper) for stars within $\Delta r_{crit}$.
If there are multiple KIC stars matching one SExtractor object,
then only the brightest star (in KIC g) is accepted, while all the others are flagged as blended.
For the brightest star, if the total computed flux of all the other stars within $\Delta r_{crit}$
contributes more than 10\% of the computed flux of the brightest star, then the brightest star is also flagged as blended.

The FWHM (by which we mean the FWHM given by SExtractor, which is computed directly from the digitized measure of photographic densities) of a star image on a plate depends on the brightness of the star and the plate properties,
as shown in Figure 2. It ranges from $10-20''$ on small plate scale plates such as the mc plates (with scale 98''/mm),
and up to $40-100''$ on large plate scale plates such as ac (606''/mm) and dnb (577''/mm).
Somewhat surprisingly, the asymptotic value at fainter magnitudes (but still $>1$mag brighter than limiting magnitude)
for the FWHM of the psf in pixel units (1 pixel for the DASCH scanner is 11microns; cf. Simcoe et al 2006) is $\sim$10 pixels, for all plate series.
However, within 0.5 mag of the limiting magnitude, the FWHM does drop to $\sim3-4$ pixels.
The astrometric error is much smaller than the FWHM,
with typical values from sub-arcsecond to a few arcseconds,
mostly depending on plate series (L10; Servillat et al. 2011).

\begin{figure}   [tb]
\epsfig{file=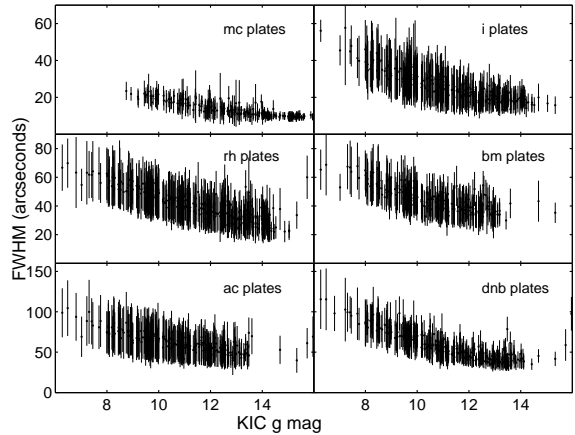, angle=0, width=\linewidth}
\caption{Image size (FWHM) of 400 randomly selected stars vs
magnitude over different plate series (6 of the $\sim$10 principal series).
For any given plate series, the solid points
present the mean value of FWHM over different plates, and the error
bars show the rms.  }
\end{figure}

\section{Filter out Dubious Detections}
\subsection{Image Profiles}
There are many kinds of defects on the plates,
including emulsion defects, scratches,
ink marks or dust particles not removed in the cleaning process for each plate,
and even airplane and satellite tracks (post 1957),
which all must be ``removed'' in the search for variables.
Ink marks from many users of the historical plates (annotations of objects they were working on,
and present on about 20\% of the Harvard plates),
and most of dust particles as well as ``smudges'',
are removed during the plate cleaning of the clear glass back side of the plate,
but other defects remain there.
Such defects must be flagged and removed for a new transient surveys or if within the astrometric error or FWHM of a known object.

Fortunately, most defects have different image profiles from real stars,
which provides a way to filter them out.
Plates are highly inhomogeneous, and
images in different parts of the plate have different profiles,
as shown in Fig. 3, which shows image parameters (from SExtractor) vs. spatial bin location on the plate;
note that this plate is a highly trailed, and images have exceptionally larger ellipticities than on average plates.
Therefore, we divided each plate into $N\times N$ bins to do the filtering, with $N=min(10,\sqrt{n/1000})$,
where $n$ is the total number of objects which are detected and matched with catalog stars.
Objects at edges (within 5\% of the edges of the plate width and height)
and corners (beyond 85\% of the distance from the plate center to one of the corners)
are excluded in the analysis to avoid contaminations,
as they are more often distorted.
We used SExtractor parameters as profile measurements,
including ellipticity, position angle ($\theta$), instrumental isophotal magnitude ($MAG\_ISO$),
FWHM (in arcsec), SExtractor FLUX$_{\rm{MAX}}$ with background subtracted (which are computed directly from the digitized measure of photographic densities),
and isophotal area above the analysis threshold (ISO0, in units of pixel$^2$).

\begin{figure}   [tb]
\epsscale{.99} \plotone{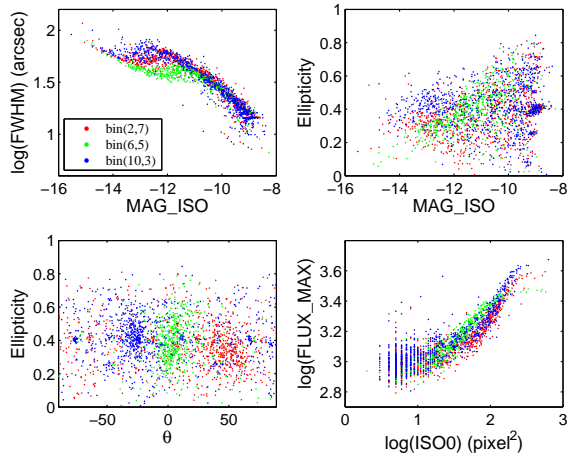}
\caption{SExtractor parameter distributions for stars in different local bins of
plate bm 27.}
\end{figure}

\emph{Filters} are designed to constrain and identify defects by selection within four regions of the available parameter space of SExtractor parameters, as shown in Figure 4.
The filters are evaluated for each spatial bin on each plate as described above.
\begin{enumerate}
\item log(FWHM) vs instrumental magnitude, as shown in the upper-left panel of Figure 4. Objects are divided into $\sim30$ instrumental magnitude bins.
In each bin, median and rms values of log(FWHM) are derived after three iterations of $3\sigma$ clipping.
Objects beyond $3\sigma$ are flagged as defects, where $\sigma$ is defined as the rms of the value in a given bin after clipping.
The threshold ($3\sigma$ here; it could be different for different parameter space, for example, $2.5\sigma$ threshold is adopted for ellipticity vs position angle)
is chosen after extensive experimenting, to minimize false positives (defects not filtered out) and wrong negatives (true images filtered out).
\item Ellipticity vs instrumental magnitude, as shown in the upper-right panel of Figure 4. Objects are divided into $\sim30$ instrumental magnitude bins.
In each bin, median and rms values of ellipticity are derived after three iterations of $3\sigma$ clipping.
Objects beyond $3\sigma$ are flagged as defects.
\item Ellipticity vs position angle, as shown in the lower-left panel of Figure 4.
Median and rms values are derived after three iterations of $2\sigma$ clipping.
Objects beyond $2.5\sigma$ in ellipticity or position angle are flagged as defects.
\item log(SExtractor FLUX$_{\rm{MAX}}$) vs log(ISO0), as shown in the lower-right panel of Figure 4. Objects are divided into $\sim30$ bins in both log(SExtractor FLUX$_{\rm{MAX}}$) and log(ISO0).
In each bin, median and rms values are derived after two iterations of $2\sigma$ clipping and one iteration of $3\sigma$ clipping.
Objects beyond $3\sigma$ are flagged as defects.
\end{enumerate}
Only good matches, i.e. images matched with KIC stars within $2\sigma$ astrometric uncertainty and therefore are most likely real objects,
as shown in blue dots in Figure 4, are used in the above calculation.
The resulting thresholds are shown in black lines, and objects beyond the black lines are considered as defects.
Red open circles are images $>4\sigma$ away from the nearest KIC stars, which are most likely defects.
As shown in Figure 4, most un-matched objects, which are most likely defects, as shown in red open circles,
are located at different regions in at least one or more parameter spaces from real star images,
and are correctly recognized.

\begin{figure}  [tb]
\epsscale{.99} \plotone{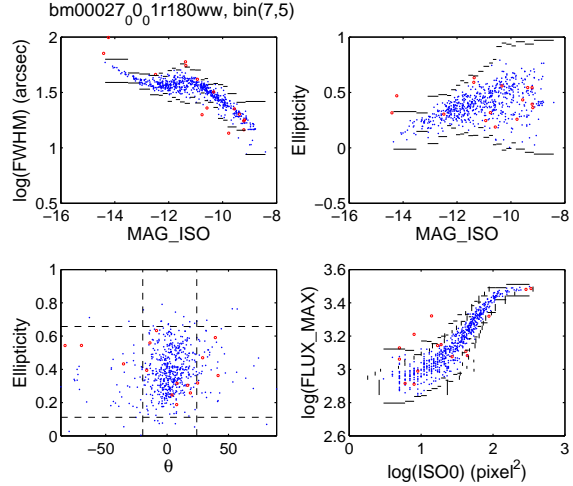}
\caption{SExtractor image profile parameter distributions for defect filter in bin (7,5) on plate bm 27.
Upper left: FWHM vs instrumental magnitude;
Upper right: Ellipticity vs instrumental magnitude;
Lower left: Ellipticity vs position angle;
Lower right: Flux$_{max}$ vs isophotal area above the analysis threshold.
Blue dots are images matched with KIC stars within $2\sigma$ astrometric uncertainty,
which are most likely real objects.
Red open circles are images $>4\sigma$ away from the nearest KIC stars, which are most likely defects.
The thresholds derived by the parameter distributions of good matches (blue dots) are shown in black lines,
and objects beyond the black lines are considered as defects and are dubious.}
\end{figure}

Both real stars and defects have wide
distributions in parameter phase spaces, and the ideal goal that all
defects are filtered out while all real stars are kept is
impossible. Tighter constraints lead to less contamination from defects,
while more real stars are lost;
looser constraints yield more real stars, but with higher level of contamination from defects.
After extensive experimenting, we chose the above $2.5-3\sigma$ cut-off to balance, and the resulted ratios are
shown in Figure 5. The objects identified as defects are flagged,
but are still kept in the pipeline so these can be plotted in light curves and included in analysis if desired.

\begin{figure}  [tb]
\epsscale{1} \plotone{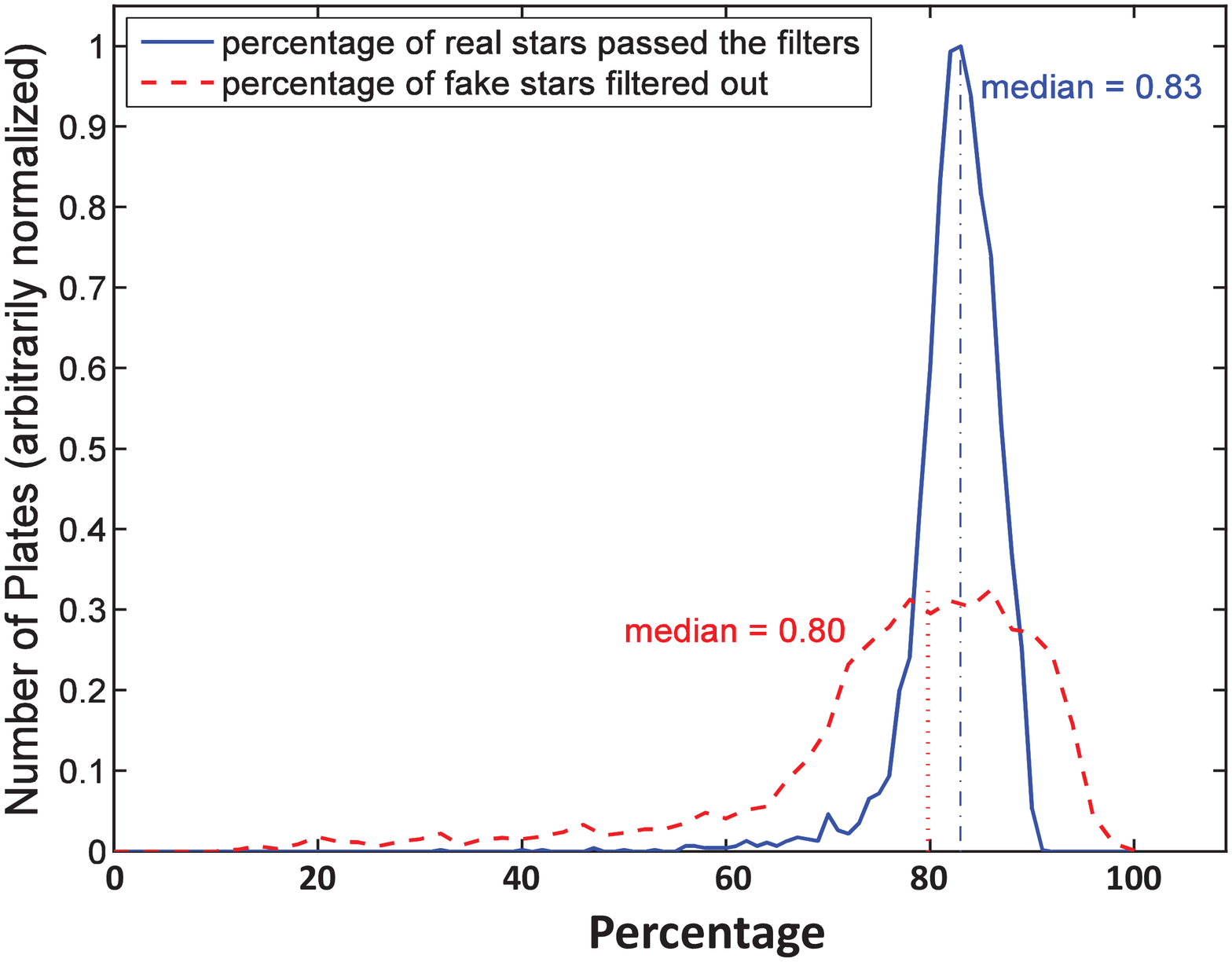}
\caption{Efficiency of defect filters (all 4 filters combined). They are based on the
SExtractor parameters characterizing the image profiles. The blue
solid line shows the percentage distribution of real stars, i.e., images
matched with KIC objects within $2\sigma$ astrometric uncertainty, that passed the
filters; its median value, i.e. 0.83, is marked by the blue dash-dotted line.
The red dashed line shows the percentage distribution of fake
stars, i.e., images not matched with KIC objects thus most likely defects,
that failed the filters; its median value, i.e. 0.80, is marked by the red dotted line.
The above results are based on 3735 plates in the Kepler field.}
\end{figure}

\subsection{\textit{drad} filter}
Since we started matching using a search radius ($FWHM/2 + $ astrometric error) much larger than our
astrometric error, we might include wrong matches, mostly real
stars matched with noise or defects. We define \textit{drad} as the
difference between the SExtractor image position and the input catalog (KIC, APASS, or GSC catalogs)
position.
The positions are adjusted for the proper motions using the UCAC4 catalog (Zacharias et al. 2013).
We first find the median $|drad|$ in the local bin,
which is \emph{initially} defined as $50\times50$ bins on each plate
but is then re-defined to have sufficient stars ($\geq20$) in each bin to derive sufficiently small uncertainties on $\sigma_{drad}$,
the rms dispersion of the \textit{drad} values in this bin.
Thus $\sigma_{drad}$ is a measure of the local astrometry uncertainty.
Stars with \textit{drad} $> 3 \sigma_{drad} + 2$ pixels are flagged as wrong matches.
The typical values of $3 \sigma_{drad} + 2$ pixels are on the order of $5-15$ arcseconds, depending on plate series.

\section{Improved local calibration using neighbors with similar magnitudes}
After defect filtering, we divide each plate into 9
annular bins (first 7 of equal area; last 2 to cover plate corners and outermost edge; see L10),
in order to proceed with photometry calibrations that
will naturally take into account radial vignetting effects from the original telescope.
In each annular bin, we derive the effective color response by minimizing the scatter
in the `effective' catalog magnitude (defined by catalog magnitude and color) vs instrumental magnitude plane,
and fit the calibration curve for each bin (L10).
After the annular calibrations,
to correct the localized spatial variations in plate sensitivity,
a first-pass local calibration is performed in $50\times50$ ``local bins'',
where a clipped median magnitude residual (plate magnitude - catalog magnitude)
is computed for each bin, and applied to each star's magnitude (L10).
As there are many more faint stars in each bin which dominate the median residual,
the first-pass local calibration accounts well for the spatial variations of the fainter stars,
but not well for bright stars, especially the ones with $B<10-11$ mag, as shown in Figure 6.
In the two brighter bins (upper panels in Figure 6), there are a large linear gradients in the two brighter bins,
with positive magnitude residuals in the upper-left corner
and negative magnitude residuals in the lower-right corner;
while in the two fainter bins (lower panels in Figure 6), the gradients have the opposite trends,
and with smaller variations.
Such residual gradients are due to the non-axisymmetric components of plate inhomogeneity,
which could not be removed in the annular calibrations.

\begin{figure}  [tb]
\epsscale{1} \plotone{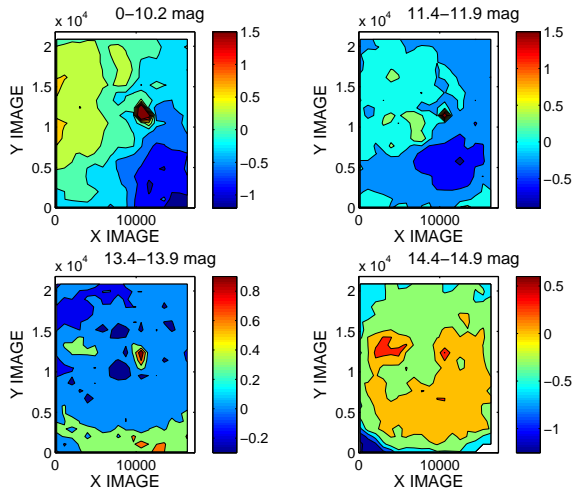}
\caption{Magnitude residual (plate magnitude - catalog magnitude) contour of plate rb07136 for stars in different magnitude bins.
The X and Y axis are in units of pixels.
Different colors are for different contour levels, as shown in the colorbars (in unit of mag).}
\end{figure}

To further improve the photometry, we perform a second-pass local calibration, using neighbor stars with similar magnitudes.
We first divide stars in magnitude bins, requiring a minimum number of stars in each magnitude bin of 3000, and a minimum bin size of 0.5 mag.
For each magnitude bin, we then divide the plate into $\sim400$ quasi-square spatial bins, for example, $17\times23$ spatial bins for a $8\times11$ inch plate.
If the number of stars in a magnitude-spatial bin is less than 20, we expand the bin until it contains at least 20 stars.
We then calculate the clipped median magnitude residual, and apply the residual correction to the magnitude of each star in the bin.
Figure 7 shows the magnitude residual correction vs magnitude we derived for 9 example spatial bins on plate rb07136.
The median magnitude residual depends on both the spatial location, which reflects the inhomogeneity of the plate,
and the magnitudes of stars, which suggests that fitting a dmag vs catalog mag to a sequence of stars around the given star would further improve the photometry.
In principle, this is similar to what people do when examining the plates using eyepieces,
i.e. comparing the image of the target object with a sequence of neighbor stars with similar magnitudes.

\begin{figure}  [tb]
\epsscale{1} \plotone{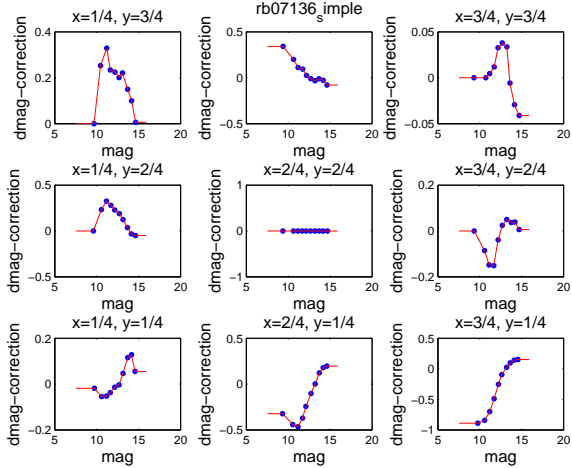}
\caption{Magnitude residual correction vs magnitude of 9 example spatial bins on plate rh07136.
$X=1/4, Y=1/4$ means the spatial bin locates at $1/4$ of both X and Y axis,
$X=2/4, Y=2/4$ means the spatial bin locates at the center of the plate, and so on.
Positive dmag correction means this must be subtracted to match the catalog value (i.e. stars are being measured as too dim).}
\end{figure}

\section{DASCH Photometry Performance and Sensitivities for Variables}
There are two measures of DASCH photometry performance.
The first is light curve rms.
As most stars are not variable at the $\sim0.1$ mag level, the light curve rms is dominated by photometric uncertainty,
and the median light curve rms of all stars corresponds to the average uncertainty.
The second is the number of unrecognized bad measurements, which ultimately constrains our ability to find real variables.
As shown in Figure 8, the second-pass magnitude-dependent local calibration
improves the median light curve rms (upper panel),
and greatly reduced the number of bad measurements by one order of magnitude (lower panel).
Here we use ``nburst3'', defined as the number of points in a given light curve of a star that are $\geq0.4$mag brighter
than the median magnitude, as a measure of the number of ``bad'' points.
In the next section, we describe how real variables are distinguished from this ``nburst3'' measure of the number of bad points.
The ``peak'' at mag $\sim12-13$ is due to this being the approximate limitng magnitude of the ``patrol plates'' (i.e. series AC, etc.);
the reduced rms at mag $\sim15$ is due to the contribution of the finer scale plates (MC, MF, A seriese) which reach these fainter magnitudes.

We have also compared the photometry performance using different input catalogs for calibration, as shown in Figure 9.
Four catalogs are used: the KIC, APASS, GSC, and an experimental catalog built using APASS B magnitudes and GSC B-V colors.
For stars brighter than 12th mag, KIC and APASS yield better results than GSC, due to more accurate photometry in these two catalogs.
However for mag $<\sim8.5$, the APASS photometry is not available and so stars are ``lost'' whereas GSC incorporates the Tycho catalog.
For stars fainter than 12th mag, the results are comparable, suggesting that intrinsic errors in the plates dominate.
Note that the experimental catalog with APASS B magnitudes and GSC $B-V$ colors has larger light curve rms than the real APASS catalog,
showing that the \emph{much better} $B-V$ color in APASS than GSC helps our photometry calibration.
Note that even using the GSC catalog, our typical photometric uncertainty, as measured by light curve rms,
is about $0.10-0.13$ mag for stars of $8-16$ mag.
Compared with the photometric uncertainty of $0.15-0.20$ mag we got in our initial photometric pipeline in 2010 (see Figure 13 in Laycock et al. 2010),
this is a big improvement, resulting from our many developments in astrometry, photometry, and defect filtering.

\begin{figure}  [tb]
\epsscale{1} \plotone{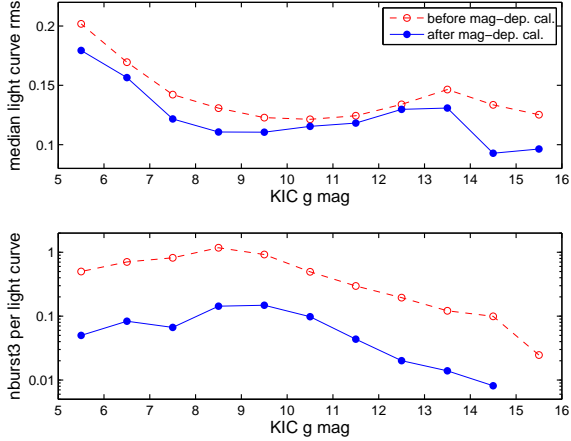}
\caption{Upper panel: median light curve rms vs KIC g mag in 1 mag bins;
Only objects with at least 10 good measurements are included.
Lower panel: number of nburst3 points per light curve vs KIC gmag,
where nburst3 points are defined as $\geq0.4$ mag brighter than median mag of a star.
The results before magnitude-dependent local calibration are shown in red open circles with dashed lines,
and the results after magnitude-dependent local calibration are shown in blue filled circles with solid lines.}
\end{figure}

\begin{figure}  [tb]
\epsscale{1} \plotone{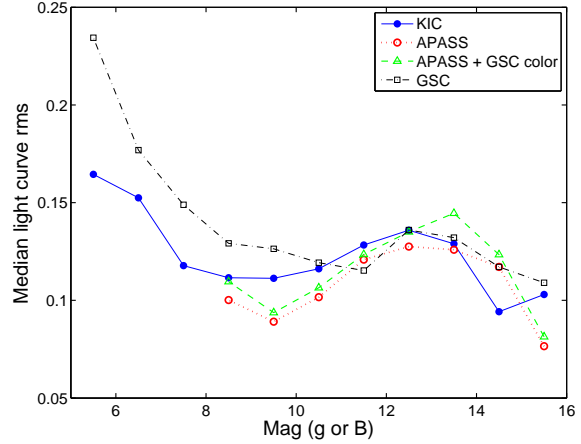}
\caption{Comparison of median light curve rms vs g or B mag, using different input catalogs for calibration.
Only stars with at least 10 good measurements are used.
Stars are divided into 1 magnitude bins.
In each magnitude bin, we calculate the DASCH light curve rms value for each star in the bin,
and take the median of all the light curve rms values as ``Median light curve rms".}
\end{figure}

\section{Statistical Measurements for Defining Variable Search}

We have developed a set of statistical measurements for the purpose of selecting variable stars.
We first define `good' points
($AFLAGS==0$ in the DASCH database; see the DASCH website\footnote{http://hea-www.harvard.edu/DASCH/database.php})
for the definitions of flags,
excluding the following cases:

\begin{enumerate}
\item Blended images either flagged by SExtractor or our pipeline (see section 3);
\item Images flagged as defects or with a large astrometric position rms (see section 4);
\item Images on multiple exposure or or ``Pickering wedge'' (for calibration) plates -- see Los et al (2011);
\item Images in those local bins with anomalously large photometry or astrometry errors, and thus likely dubious;
\item Images on plates with uncertain date, or within 23.5 degrees of horizon which are likely have wrong observing dates recorded;
\item Images which are too bright (typically $B <8-9$) to be calibrated, due to the lack of calibration stars;
\item Images matched with galaxies given by the input catalog (this ``galaxy flag'' does not eliminate them as candidate variables).
\end{enumerate}

Using these `good points' (for selecting stellar variables),
we then calculate the statistical measurements of each light curve.
The measurements used for selecting variables are listed in the ``summary table'' shown in Table 2,
and are described in the following paragraphs.
Description of other parameters in the summary table can be found at DASCH website\footnote{http://hea-www.harvard.edu/DASCH/database.php}.

Two amplitude measurements are calculated. The first one is called $range\_local$,
which is the difference between the brightest and the faintest points, minus the sum of their errors.
The second one is called $range\_local2$, which is similar to $range\_local$,
but after removing the brightest and the faintest points.
Light curve rms is calculated after $5\sigma$ clipping,
and is called $lightcurverms1$ in our database.

We also de-trend the light curves by smoothing the light curves using four different combinations of spans and methods,
and then calculate the rms of the residuals, to look for variables with trends,
which have residual rms significantly smaller than their light curve rms.
Methods used for de-trending are listed in Table 2.
The first two ($lightcurverms2$ and $lightcurverms3$) are sensitive
to variables with slow variations over years to decades,
while the other two  ($lightcurverms4$ and $lightcurverms5$)
are sensitive to variables with more than $10-15$ adjacent points in the trend.

We use a set of different thresholds to define `outburst', `dip', and `dev' (deviation) points.
If there are multiple adjacent points in a `outburst' or `dip' in a light curve,
then it is more likely to be a true variable.
To quantify this, we define parameters ($adjacentburstdip$, $adjacentburstdip2$, and $adjacentburstdip3$)
to measure the number of adjacent points in a light curve.

To further exclude possible dubious variables which were not caught by our blending and defect filters,
we calculate a set of correlation measurements,
including the correlation coefficients between magnitude and ra, dec, and limiting magnitude.
We also calculate the difference between the median DASCH magnitude of 20 deepest plates,
and 20 shallowest plates.
Blends and noise/defects trends to have significantly higher correlations between magnitude and other properties (ra, dec and limiting magnitude).

When using the statistical measurements in the summary table for a given object,
it is always recommended to compare the measurement values with its neighbor stars with similar magnitudes.
DASCH is an inhomogeneous survey:
Stars in different regions of the sky have different plate coverage;
At a given region of the sky, there are multi-series plates with different plate scales and limiting magnitudes;
On a given plate, the DASCH magnitude uncertainty of an object is a function of both the magnitude of the object, and the location of the object.
The significance of variability or dubious correlation of a star,
can not be derived from the statistical measurements of the star alone.
The properties of DASCH plates covering it must be taken into account,
for which the best and easiest way is to compare the statistical measurements of the star with its neighbors with similar magnitudes.

\section{Summary}
In this paper, we describe our photometry and variable identification software that extend the original pipeline (Laycock et al. 2010),
in order to improve our science goal of finding real variables on
the 100 years of Harvard plates being digitized by DASCH.
These developments significantly improved the efficiency of removing dubious measurements,
as well as reduced the overall photometric uncertainty to $0.1-0.12$ mag level.
DASCH variable identifications can now start with the summary table (Table 2),
by selecting candidates with excess variability measurements compared with neighbor stars with similar magnitudes.
Details of an actual implementation of a variable search
using the methods described here are presented in our Kepler field variable search, reported in Tang et al (in prep.).

\acknowledgments
We thank the referee, George Djorgovski, for many helpful comments and suggestions that improved the paper.
There are many colleagues who have helped out. In particular we thank our team members
Alison Doane, Bob Simcoe, Jaime Pepper, David Sliski, and Silas Laycock for their work on DASCH.
We would also like to thank many volunteers who have helped digitize logbooks,
clean and scan plates (\emph{http://hea-www.harvard.edu/DASCH/team.php}).
This work was supported in part by NSF grants AST0407380 and AST0909073
and now also the \emph{Cornel and Cynthia K. Sarosdy Fund for DASCH}.
This research has made use of the APASS database, located at the AAVSO web site.
Funding for APASS has been provided by the Robert Martin Ayers Sciences Fund.\\

\begin{table*} [tb]
\caption{Plate series characteristics and numbers for 3735 digitized plates in the Kepler field.} \centering
\tabcolsep 3pt
\begin{tabular}{lrcccc}
\tableline
Series	&	Number	&	Lens aperture	&	FOV	&	Median limiting & RMS of limiting 	\\
	&	of plates	&	(inches)	&	($^{\circ}$)	&	magnitude &	magnitude		\\
\tableline
ac	&	1578	&	1.5	&	$34\times43$	&	12.1	&	0.9	\\
i	&	690	&	8	&	$9\times12$	&	13.8	&	0.9	\\
dnb	&	430	&	1.6	&	$33\times41$	&	14.2	&	0.7	\\
rh	&	342	&	3	&	$22\times28$	&	13.4	&	1.1	\\
mc	&	321	&	16	&	$6\times7$	&	17.0	&	1.2	\\
bm	&	96	&	1.5	&	$22\times27$	&	13.6	&	0.9	\\
a\tablenotemark{1}	&	75	&	24	&	$3\times4$; $6\times7$	&	14.0	&	1.1	\\
ay	&	63	&	2.6	&	$39\times49$	&	12.5	&	0.8	\\
dnr	&	47	&	1.6	&	$33\times41$	&	11.7	&	0.7	\\
ca	&	40	&	2.5	&	$34\times42$	&	12.3	&	0.8	\\
ma	&	19	&	12	&	$5\times7$	&	16.0	& 1.5	\\
dny	&	12	&	1.6	&	$33\times41$	&	13.8	&	1.0	\\
mb	&	8	&	3 to 6	&	$22\times28$	&	13.5	&	0.7	\\
aco	&	5	&	1	&	$35\times43$	&	12.6	&	1.1	\\
am	&	4	&	1	&	$34\times43$	&	12.5	&	0.8	\\
ax	&	4	&	3	&	$39\times49$	&	12.9	&	0.7	\\
me	&	1	&	4	&	$11\times13$	&	11.4	&	0.2	\\
\tableline
\end{tabular}
\tablenotetext{1}
{Note that the Bruce telescope for `a' series plates was moved from Cambridge, MA (latitude $+42^\circ$)
to Arequipa, Peru (latitude $-16^\circ$) in 1896, and later to
Bloemfontein, South Africa (latitude $-29^\circ$) in 1929.
At $Dec\sim+45^\circ$, the Kepler field was very low in the sky when the telescope was at Arequipa ($1896-1928$),
and even lower when the telescope was at Bloemfontein (after 1929).
As a result, most (73 out of 75) `a' series plates processed in the Kepler field were taken before 1910,
with much poorer emulsions compared with later ones, and are much shallower than `a' plates on average, which have typical limiting magnitude of $\sim17-18$ mag.}
\end{table*}

\begin{table*}[tb]
\caption{Statistical measurements in the summary table} \centering \leavevmode
\begin{minipage}{\textwidth}
\tabcolsep 4pt \small
\begin{tabular}{ll}
\tableline Parameter  & Description \\
\tableline
Group 1. & Light curve amplitude and rms: \\
\tableline
$range\_local$ & difference between the brightest and the faintest points, minus the sum of their errors  \\
$range\_local2$ & similar to range\_local, but after removing the brightest and the faintest points. \\
$lightcurverms1$ & light curve rms after 4 iterations of $5\sigma$ clipping \\
\tableline
Group 2. & rms of light curve residuals after de-trending:  \\
\tableline
$lightcurverms2$ & de-trended using smooth(x,y,0.4, `lowess') \\
$lightcurverms3$ & de-trended using smooth(y,0.8, `lowess') \\
$lightcurverms4$ & de-trended using smooth(y,10, `sgolay')  \\
$lightcurverms5$ & de-trended using smooth(y,15, `loess') \\
\tableline
Group 3. & Number of outburst and dip points: \\
\tableline
$nburst$ & number of points $\geq0.8$ mag brighter than the median value \\
$nburst2$ & number of points $\geq0.5$ mag brighter than the median value \\
$nburst3$ & number of points $\geq0.4$ mag brighter than the median value \\
$nburst4$ & number of points $\geq3\sigma$ brighter than the median value, \\
        & where $\sigma$ is the median value of photometry uncertainty in the light curve\\
$ndip$ & number of points $\geq0.8$ mag fainter than the median value \\
$ndip2$ & number of points $\geq0.5$ mag fainter than the median value \\
$ndip3$ & number of points $\geq0.4$ mag fainter than the median value \\
$ndip4$ & number of points $\geq3\sigma$ fainter than the median value \\
$ndev2$ & number of points $\geq2\sigma$ brighter or fainter than the median value \\
$ndev3$ & number of points $\geq3\sigma$ brighter or fainter than the median value \\
\tableline
Group 4. & Adjacent points in `burst' or `dip':  \\
\tableline
$adjacentburstdip$ & a measure of the number of adjacent nburst3/4 and ndip3/4 points \\
$adjacentburstdip2$ & a measure of the number of $>5$ adjacent nburst3/4 and ndip3/4 points \\
$adjacentburstdip3$ & a measure of the number of $>7$ adjacent nburst3/4 and ndip3/4 points \\
\tableline
Group 5. &Parameters used to remove dubious variables: \\
\tableline
$magvsracorr$ & correlation coefficient between light curve magnitude and ra \\
$magvsdeccorr$ & correlation coefficient between light curve magnitude and dec \\
$magvslimitingcorr$ & correlation coefficient between light curve magnitude and plate limiting mag \\
$Malmquist\_factor$ & clipped median DASCH magnitude of 20 deepest plates  \\
& $-$ clipped median DASCH magnitude of 20 shallowest plates using `good' points \\
$Malmquist\_factorB$ & similar to Malmquist\_factor but also includes defects, low altitude, \\
& uncertain date and second quality plates\\
\tableline
\end{tabular}
\end{minipage}
\end{table*}

\end{document}